\documentclass[sigconf]{acmart}

\makeatletter
\def\subsubsection{\setlength\parindent{10pt}\@startsection{subsubsection}{3}%
  \z@{.5\linespacing\@plus.7\linespacing}{.1\linespacing}%
  {\normalfont\itshape}}
\makeatother

\usepackage{booktabs} 
\usepackage{multirow}
\usepackage{amssymb}
\usepackage{amsmath}
\usepackage{caption}
\usepackage{multicol}
\usepackage{graphicx}
\usepackage{subfig}
\usepackage{enumitem}
\usepackage{siunitx}

\setcopyright{rightsretained}

\def\:{\hskip0pt} 

\theoremstyle{definition}
\newtheorem{definition}{Definition}

\setcopyright{acmlicensed}
\acmConference[LND4IR '18]{ACM SIGIR 2018 Workshop on Learning from Limited or Noisy Data}{July 12, 2018}{Ann Arbor, Michigan, USA} 
\acmYear{2018}
\copyrightyear{2018}
\acmDOI{}
\acmISBN{}
\acmPrice{}

\author{Hamed Zamani}
\affiliation{%
  \institution{Center for Intelligent Information Retrieval}
  \institution{University of Massachusetts Amherst}
  \city{Amherst, MA 01003}
}
\email{zamani@cs.umass.edu}

\author{W. Bruce Croft}
\affiliation{%
  \institution{Center for Intelligent Information Retrieval}
  \institution{University of Massachusetts Amherst}
  \city{Amherst, MA 01003}
}
\email{croft@cs.umass.edu}

\begin{document}
\title{Towards Theoretical Understanding of\\Weak Supervision for Information Retrieval}

\begin{abstract}
Neural network approaches have recently shown to be effective in several information retrieval (IR) tasks. However, neural approaches often require large volumes of training data to perform effectively, which is not always available. To mitigate the shortage of labeled data, training neural IR models with weak supervision has been recently proposed and received considerable attention in the literature. In weak supervision, an existing model automatically generates labels for a large set of unlabeled data, and a machine learning model is further trained on the generated ``weak'' data. Surprisingly, it has been shown in prior art that the trained neural model can outperform the weak labeler by a significant margin. Although these obtained improvements have been intuitively justified in previous work, the literature still lacks theoretical justification for the observed empirical findings. In this position paper, we propose to theoretically study weak supervision, in particular for IR tasks, e.g., learning to rank. We briefly review a set of our recent theoretical findings that shed light on learning from weakly supervised data, and provide guidelines on how train learning to rank models with weak supervision.
\end{abstract}


\maketitle

\section{Introduction}
\label{sec:intro}
Neural network models have recently shown promising results in a number of information retrieval (IR) tasks, including ad-hoc retrieval \cite{Guo:2016}, answer sentence retrieval~\cite{Yang:2016}, and context-aware ranking \cite{Zamani:2017:WWW}. Neural approaches often require a large volume of training data to perform effectively. Although large-scale relevance signals, e.g., clickthrough data, are available for a few IR tasks, e.g., web search, this data is not available for many real-world problems and domains. Moreover, academia and smaller companies also suffer from lack of access to large-scale labeled data or implicit user feedback. This is critical for fields, such as information retrieval, that have been developed based on extensive and accurate evaluations. The aforementioned limitations call for developing effective learning approaches to mitigate the shortage of training data. In this line of research, weak supervision has been proposed to train neural models for information retrieval tasks, such as learning to rank documents in the context of ad-hoc retrieval \cite{Dehghani:2017} and learning relevance-based word embedding \cite{Zamani:2017:SIGIR}. The substantial improvements achieved by weakly supervised IR models have recently attracted the attention of the IR community~\cite{Cheng:2017,Chaidaroon:2018,Dehghani:2017:arxiv,Nie:2018,Voskarides:2018,Zamani:2018:SIGIR,Zhen:2017}. Although the obtained improvements have been intuitively well justified in previous work \cite{Dehghani:2017,Zamani:2017:SIGIR}, to the best of our knowledge, no theoretical justification has been proposed to support the empirical findings from weak supervision in information retrieval.

To close the gap between theory and practice, thorough theoretical analysis is required. We believe that theoretical understanding of learning from weakly supervised data could potentially provide guidelines on how to design and train effective models using weak supervision. In this position paper, we review our recent theoretical findings that shed light on learning from weakly supervised data for information retrieval. Refer to \cite{Zamani:2018:ICTIR} for more details.

\vspace{-0.2cm}
\section{Weak Supervision for IR}

In this section, we formalize learning from weakly supervised data and further briefly describe the different ways that we can benefit from this learning strategy in the context of information retrieval.\footnote{Note that weak supervision is an established sub-field of machine learning. In this paper, we focus on weak supervision for information retrieval.}

In typical supervised learning problems, we are given a training set $\mathcal{T} = \{(\mathbf{x}_1, \mathbf{y}_1), (\mathbf{x}_2, \mathbf{y}_2), \cdots, (\mathbf{x}_m, \mathbf{y}_m)\}$ with $m$ elements, where $\textbf{x}_i$ is feature vector(s) for the $i$\textsuperscript{th} training instance and $\textbf{y}_i$ denotes the corresponding true label(s). In classification and regression, $\textbf{x}_i$ is a vector containing the features for the corresponding item; in contrast, in learning to rank, $\textbf{x}_i$ is a list of $n$ feature vectors representing $n$ items in a rank list. In weak supervision, however, the true labels (i.e., $\textbf{y}_i$s) are unknown, which is similar to typical unsupervised learning problems. Weak supervision assumes that a pseudo labeler (or ``weak'' labeler) is available that can generate labels for all the feature vectors in $\mathcal{T}$. This results in a weak supervision training set $\widehat{\mathcal{T}} = \{(\mathbf{x}_1, \widehat{\mathbf{y}}_1), (\mathbf{x}_2, \widehat{\mathbf{y}}_2), \cdots, (\mathbf{x}_m, \widehat{\mathbf{y}}_m)\}$, where the labels (i.e., $\widehat{\mathbf{y}}_i$s) are generated using a weak labeler $M$. Learning a model $\mathcal{M}$ from $\widehat{\mathcal{T}}$ is called weak supervision. If $M$ is an unsupervised model, then $\mathcal{M}$ is also unsupervised.

Weak supervision can be used with one of the following goals:
\begin{itemize}[leftmargin=*]
    \item \textbf{Improving effectiveness:} \citet{Dehghani:2017} showed that a simple pairwise neural ranking model trained on the labels generated by BM25 as the weak labeler can outperform the weak labeler with a significant margin. More recently, \citet{Zamani:2018:SIGIR} studied the problem of learning from multiple weak supervision labels to achieve state-of-the-art results on the query performance prediction task. These studies suggest that weak supervision can be used to improve the effectiveness.
    
    \item \textbf{Bypassing the need to external resources:} \citet{Zamani:2017:SIGIR} proposed to learn relevance-based word embedding based on the relevance distributions generated by the \citeauthor{Lavrenko:2001}'s relevance models \cite{Lavrenko:2001}. They shows that not only the learned word embedding vectors are better representations compared to the general-purpose word embeddings, e.g., word2vec \cite{Mikolov:2013}, but also can perform on par with the relevance models for the query expansion task without the need to the top retrieved documents (i.e., pseudo-relevant documents) at the testing time. This shows that the huge number of parameters in neural networks are capable of memorizing useful information that can be captured from external resources and weak supervision can be employed as an approach for getting rid of these resources at the testing time. 
    
    \item \textbf{Improving efficiency:} Recently, \citet{Cohen:2018:SIGIR} showed that expensive regression forest learning to rank models, e.g., LamdaMART, can be replaced by simple feed-forward networks. The network is trained with a weak supervision setting, where the learning to rank model plays the role of weak labeler. The authors demonstrated that up to 10x (on CPU) and 100-1000x (on GPU) speeds up can be obtained compared to state-of-the-art implementation of regression forest learning to rank models, with no measurable loss in effectiveness. This recent work suggests that weak supervision can be used to obtain more efficient models.
    
    \item \textbf{Learning from private data:} \citet{Dehghani:2017:neuir} proposed to share a model trained on sensitive private data, instead of sharing the data itself. Although this should be done in caution due to various membership attacks, see \cite{Shokri:2017}, the shared model can be used as the weak labeler.
    
\end{itemize}

\vspace{-0.2cm}
\section{Our Theoretical Findings}
We believe that the following theoretical questions are research-worthy and answering them sheds light on learning from weak supervision.

\begin{itemize}[leftmargin=*]
    \item Why and how a weakly supervised model can outperform the weak labeler?
    \item What properties should the weakly supervised models have to perform effectively?
\end{itemize}

Our recent work \cite{Zamani:2018:ICTIR} studies weak supervision for information retrieval with a focus on learning to rank and models weak supervision as a noisy channel that introduces some noise to the true labels. Motivated by the symmetry condition defined for classification \cite{Ghosh:2015}, we define symmetric ranking loss function as follows.

\begin{definition}{[Symmetric Ranking Loss Function]}
A ranking loss function $\mathcal{L}(\cdot, \cdot)$ is symmetric, if it satisfies the following constraint:
\begin{equation*}
    \sum_{\mathbf{y} \in \mathcal{Y}}{\mathcal{L}(\mathcal{M}(\mathbf{x}), \mathbf{y})} = c \quad \forall \mathbf{x}, \forall \mathcal{M}
\end{equation*}
where $c$ is a constant number and $\mathcal{Y}$ is a finite and discrete output space. In case of binary relevance judgments, the output space $\mathcal{Y}$ is $\{0, 1\}^n$ for a rank list of $n$ items.

\label{def:symmetric}
\end{definition}

Based on the risk minimization framework, we proved that symmetric ranking loss functions are noise tolerant under the uniform weak supervision noisy channel assumption. On the other hand, with a non-uniformity assumption, we find an upper bound for the risk function of the model trained on the weak supervision data. Our theorems provide insights into how and why training models on weakly supervised data can perform well and even outperform the weak labeler. They also introduce some guidelines on what loss functions to use while training on weakly supervised data. We also studied how learning from multiple weak supervision signals can improve the performance and found an information theoretic lower bound for the number of independent weak labelers required to guarantee an arbitrary maximum error probability of $\epsilon$. More information can be found in \cite{Zamani:2018:ICTIR}.

\vspace{-0.2cm}
\section{Conclusions and Future Work}
\label{sec:conclusion}
In this position paper, we provided a brief overview of weak supervision for information retrieval and how one can benefit from this learning strategy. We reviewed a set of our recent theoretical findings towards the theoretical understanding of weakly supervised IR models. In this paper, we only focus on studying the effectiveness of weakly supervised models. Since weak supervision requires large volumes of training data, we intend to theoretically study the efficiency of weakly supervised models in terms of training time. Reducing their training time is an interesting direction.


\small{
\noindent \textbf{Acknowledgements.}
This work was supported in part by the Center for Intelligent Information Retrieval. Any opinions, findings and conclusions or recommendations expressed in this material are those of the authors and do not necessarily reflect those of the sponsor.
}



\end{document}